\documentstyle[11pt]{article}
 
\def\pa{\partial}
\def\k{\kappa} 
\def\g{\gamma} \def\G{\Gamma}
\def\a{\alpha} 
\def\b{\beta} 
\def\d{\delta} \def\D{\Delta}
\def\e{\epsilon} 
 
\def\k{\kappa}
\def\l{\lambda} \def\L{\Lambda}
\def\m{\mu} 
\def\n{\nu}

\def\s{\sigma}

\def\o{\omega} 
\def\be{\begin{equation}}
\def\ee{\end{equation}}
\title{Uniqueness of D=11 Supergravity}
\author{S. Deser\\ Department of Physics\\
Brandeis University\\
Waltham, MA 02254, USA}

\begin{document}

\maketitle

\begin{abstract}
We study the extent to which D=11 supergravity can be
deformed and show in two very different ways
that, unlike lower D versions, it forbids an extension 
with cosmological constant.  Some speculations about
other invariants are made, in connection with the 
possible counterterms of the theory.
\end{abstract}

%\vspace{.1in}
\vfill

\begin{flushright}
ULB-TH-97/07\\
BRX-TH-424\\
\end{flushright}
\vfill

It is a pleasure to report here on work done jointly
with K.\ Bautier, M.\ Henneaux, and D.\ Seminara, one of
whom is closely connected with this Center and all of 
whom I thank for their contribution to this paper.  The
completed aspects of our research have also just appeared in
print \cite{00a}.

Anyone studying supergravities cannot fail to marvel at how
the interplay of Lorentz invariance, Clifford algebra and 
gauge field properties conspire to limit their
dimensionality.  In particular not only is
D=11 maximal if gravity is to remain the highest spin
of the supermultiplet, but
only one configuration of fields, the N=1 graviton
plus vector-spinor and 3-form potential combination
is permitted\footnote{For a quick counting,
recall that graviton excitations are described
by transverse traceless spatial components $g^{TT}_{ij}$,
hence there are 
(D-2)(D-1)/2 -1 = D(D-3)/2; the spatial three-form
$A^T_{ijk}$ are also
transverse hence (D-2)(D-3)(D-4)/3!,  while
the vector-spinors have (D-3) 2$^{[D/2]-1}$ 
excitations.}\cite{001}. There is some ``fine print" 
involved as well.  For example, I 
will be talking solely about actions whose gravitational
component is Einstein, rather than say Chern--Simons --
where very recent work at this Center \cite{00c}
has revealed other supersymmetric D=11 
possibilities.  Beyond
D=11, one would require the appearance of spin $>$2 fields
and/or more than one graviton, both of which are known
\cite{00d,00e} to lead to inconsistencies.

Despite the magic of D=11, the theory was for many years
neglected (if never forgotten -- as an inspiration for
KK descents to lower dimensions, if nothing else), because
superstrings had their own magic number, D=10.  Faith in 
the importance of D=11 was revived when it was seen to be
the low energy sector of M-theory unification, and
that ``dimensional enhancement" was as interesting as
dimensional reduction.  
So this is a good time to understand more
deeply just how unique a theory 
it really is -- within the framework
I have indicated, requiring
that it have an Einstein term to describe the
graviton.  Now there is one other question in
physics that is on an equally mysterious footing
as what is so special about D=11, namely why is
$\Lambda$=0 -- the
cosmological constant problem.  Early hopes that 
supersymmetry would solve it in the matter sector, 
through vacuum energy cancellations, were made to some
extent irrelevant by the perfect consistency of 
cosmological constant extensions of supergravity.  A
cosmological term (of anti deSitter type) 
$\sim -|\Lambda | \sqrt{-g}$ could be added to a
masslike term for the fermion $\sim \sqrt{|\L |} \:
\bar{\psi}_\m  \G^{\m\n} \,\psi_\n$ in a supersymmetric
way, as was first realized for D=4 
\cite{Townsend,deszum} and then
extended all the way to D=10 \cite{Roman}.
Things got rather elaborate on the way up,
with scalars for example decorating the 
cosmological term, but it was there.  At D=11,
however, there seemed to be a snag; surprisingly, there
were only a couple of papers directed at this question
at the time. To our knowledge,  
there have been two
previous approaches to  this result. One \cite{Nahm}
consists
in a classification of all graded algebras and
consideration of their highest spin representations.
Although we have not found an explicit exclusion of 
the cosmological extension
in this literature, it is undoubtly implied
there under similar assumptions.
The second 
\cite{sagnotti} considers the properties
of a putative ``minimal" graded Anti de Sitter algebra 
and shows it to be inconsistent in its simplest form. 
While one may construct generalized algebras that 
still contract to super-Poincar\'e, these can also 
be shown to fail, using for example  some results 
of \cite{Fre}. In \cite{sagnotti},
a Noether  procedure, starting from the full theory of 
\cite{001}, was also attempted; as we shall show below, 
there is an underlying cohomological 
basis for that failure.
A careful reconsideration of the problem, 
resulting in a no-go theorem is
the main result to be reported here.  To be sure, this
does not exorcise the cosmological constant problem: it can
reappear under dimensional reduction (as in fact discovered
again recently \cite{West}) as well of course 
as through supersymmetry
breaking.  Still, it should be appreciated that there is now 
one model--and a very relevant one it is--of a QFT that 
includes gravity and really excludes a $\L$ term, through
the magic of supersymmetry.  Although we have no ``deep"
physical selection rule to account for this, we can point
to the mysterious 3-form field as the immediate cause.  We
also mention that current investigations of lower 
dimensional (brane) models also have a stake in the
outcome (see eg. \cite{002}.)

We will proceed from two 
complementary starting points.  The first will
be the Noether current approach, in which we 
attempt---and fail!---to find a linearized,
``globally" supersymmetric model 
about an Anti de Sitter (AdS) background
upon which to construct a full locally
supersymmetric theory.  Since a Noether procedure
is indeed  a standard way to obtain the full 
theory, in lower dimensions,  the absence of a
starting point for it effectively  
forbids the extension. In contrast, the second
procedure will begin with the full (original)
theory of \cite{001} and attempt, using cohomology 
techniques, 
to construct---also unsuccessfully---a consistent
deformation of the model and of its transformation
rules that would include the desired fermion mass term 
plus cosmological term extensions.  In both cases, the
obstruction is due to the $4-$ (or $7-$) form field 
necessary to balance degrees of freedom.

First, we recall some general
features relevant to the linearized approach.  
It is well-known that Einstein theory with
cosmological term linearized 
about a background solution of constant curvature  
retains its gauge 
invariance and degree of freedom count, with
the necessary modification that the vielbein
field's gauge transformation is the background
covariant  $\d h^a_\m = D_\m\xi^a$.
Similarly it is also known that the free spin
3/2 field's gauge invariance in this space
is no longer
$\d \psi_\m = \pa_\m \a (x)$ or even
$D_\m \a (x)$, but rather the extended form
\cite{deszum}
\be%1
\d \psi_\m = {\cal D}_\m \, \a (x) \equiv
(D_\m + m\g_\m ) \a (x)
\ee
where  ${\cal D}_\m$  has the
property that $[{\cal D}_\m , {\cal D}_\n ] =0$
when the mass $m$ is ``tuned'' to an AdS
cosmological constant: $2 m= \sqrt{-\Lambda}$
(in $D=11$).  The modified
transformation (1)
then keeps the degree of freedom count for
$\psi_\m$ the same as in flat space, 
provided---as is needed for consistency---that
the $\psi$'s action and field equations also
involve ${\cal D}_\m$ rather than $D_\m$.
[This is of course the reason for the ``mass" 
term $m \bar{\psi}_\m \G^{\m\n} \psi_\n$
acquired by the spinor field to accompany the
cosmological one for gravity.]  Given the above
facts, the 3-form potential  $A_{\m\n\rho}$
still  balances fermi/bose degrees
of freedom here.
[For now, we keep the same field content as in
the flat limit.]
Unlike the other two fields, 
its action only involves curls and so 
it neither needs nor can accomodate any extra
terms in the background to retain its gauge
invariance and excitation count; indeed, the 
only possible quadratic addition would be a
-- true -- mass term $\sim \L \, A^2$ that would
destroy both  (there would be 120, instead of the 84 
massless, excitations).  One can
therefore expect, with reason,
that the problem will lie in
the form (rather than gravity) sector's
transformation rules.  In the AdS background,
the desired ``globally" 
supersymmetric free field starting point
involves the 
Killing spinor
%``most constant" spinor 
$\e (x)$,
%namely one that 
${\cal D}_\m \e (x) =0$,
which is unrelated to the general gravitino gauge
spinor  $\a (x)$ in (1).
[Note that we can neither use $\pa_\m \e = 0$ 
because space is curved, nor
$D_\m \e =0$ because only
${\cal D}_\m$'s commute.]  The rules are 
essentially fixed from the known flat background
ones (to which they must reduce for $\L = 0$),
\begin{eqnarray}%2
\d \, \psi_\m & = & \d_h \psi_\m  + 
\d_A \psi_\m = 
\left( \frac{1}{4} X_{\m ab} (h) \G^{ab} -
m \g^a h_{\m a} \right) \e 
 +  i/144 \:
(\G^{\a\b\g\d}~_\m - 8 \, \G^{\b\g\d}
\d^\a_\m ) \e \, F_{\a\b\g\d} \; \nonumber\\
\d \, h_{\m a} & = & -i \,  \, \bar{\e} \,
\G_a\psi_\m \hspace{.4in} 
\d \, A_{\m\n\rho} = 3/2 \, \bar{\e} \,
\G_{[\m\n} \psi_{\rho ]}.
\end{eqnarray}
The linearized connection $X(h)$ is 
derived by a linearized ``vanishing torsion"
condition $D_\m h_{\n a} + X_{\m ab}
e_\n^b - (\n\m ) = 0$; throughout, the
background vielbein is $e_{\m a}$ and its
connection is $\o_{\m ab} (e)$.  Now vary the
spinorial action $I[\psi]=-1/2\int (dx) \psi_\m
\G^{\m\a\b} {\cal D}_\a \psi_\b$ (world
$\G$ indices are totally antisymmetric and 
$\G^\m = e^\m~\!_a \g^a$ etc.).  It is
easily checked that although $[\G , {\cal D}]
\neq 0$, varying $\bar{\psi}$ and $\psi$
does yield the same contribution, and using
(2) we find
\begin{eqnarray}%3
\lefteqn{\d I[\psi ] = \d_h I[\psi ] + \d_A
I [\psi ] =} \nonumber \\
& - & i/8 \int (dx) E^{\m b} (-i\k \bar{\e} \G_a 
\psi_\m ) - i/8 \int (dx) 
[ D_\a F^{\a\m\rho\s} (\bar{\e} 
\G_{[\m\n} \psi_{\rho ]} )
+ m \bar{\psi}_\m (\Gamma^{\mu\alpha\beta\rho\sigma}
F_{\alpha\beta \rho\sigma} ) \e ] \; .
\end{eqnarray}
Here $E^{\m b} $ is the variation of the 
Einstein cosmological action linearized
about AdS.  
The form-dependent piece of (3) has a first
part that behaves similarly, namely it is 
proportional to the  form field action's variation
$D_\a F^{\a\mu\rho \s}$ (the Chern--Simons
term, being cubic, is absent at this level).
With the transformation choice (2),
the variation of the Einstein plus form actions
almost  cancels (3).  There 
remains $\bar \psi F\epsilon$, the $A-$variation of
the gravitino mass term.  What possible
deformations of the transformation rules (2) and of
the actions might cancel this unwanted term?  The
only dimensionally allowed change in (2) is a
term $ \bar{\d} \psi_\m \sim m A\!\!\!/_\m
\e$; however, it will give rise to unwanted 
gauge-variant contributions from the
$m \bar{\psi} \G\psi$ term $\sim m^2 \bar{\psi}
\G A\e$, that would in turn require a true mass
term $I_m{[A]} \sim m^2 \int (dx) A^2$ to cancel,
thereby altering the degree of freedom count.
  Indeed these two deformations,
$\bar{\d} \psi_\m$ and $I_m [A]$, are the only
ones that have nonsingular
$m\rightarrow 0$ limits.  A detailed calculation 
reveals, however, that even with these added
terms, the action's invariance cannot be preserved.
In particular, there are already variations of
the $A^2$ term that cannot be compensated. 
A completely parallel calculation starting
with a dual, 7-form, model yields precisely the
same obstruction\footnote{The $7-$ form  variant was 
originally considered by \cite{Nicolai}, who
argued that it was excluded in the non-cosmological
case, but the possibility for a cosmological 
extension was not
entirely removed; the latter was considered 
and rejected at the Noether level in \cite{sagnotti}.}: 
defining the $4-$form
dual of the $7-$form, we have the same structure
as  the  $4-$form case, up to normalizations, and
face the same non-cancellation problem; also here
a mass term is useless. 

Our second approach analyses the extension problem
in the light of the
master equation and its consistent deformations 
\cite{BH,Julia,Stasheff}; see \cite{HT} for a review
of the master equation formalism appropriate to the
subsequent cohomological considerations. 
One starts with the solution of the master equation
$(S,S) = 0$ 
\cite{HT,ZJ} for the action of an undeformed theory 
(for us that of \cite{001}).
One then tries to perturb it,
$
S \rightarrow S'=S + g \Delta S^{(1)} + g^2 
\Delta S^{(2)} + .....$,
where $g$ is the deformation parameter, in such a way 
that the
deformed $S'$ still fulfills the master equation
$
(S', S') = 0. \label{masterdefo}
$
As explained in \cite{BH} any deformation of the
action of a gauge theory and of its gauge symmetries, 
consistent in
the sense that the new gauge transformations are indeed
gauge symmetries of the new action, leads to a deformed
solution $S'$ of the master equation. Conversely, any
deformation $S'$ of the original solution $S$ of the 
master equation defines a consistent deformation 
of the original gauge invariant action and of its 
gauge symmetries.  
In particular, the antifield--independent  term in
$S'$ is the new, gauge-invariant action; the terms
linear in the antifields conjugate to the classical 
fields define the new gauge transformations \cite{BH,GW} 
while the other terms in $S'$ contain information 
about the deformation of the gauge algebra and of 
the higher-order structure functions.
To first order in 
$g$, $(S^\prime , S^\prime)=0$ implies  
$(S,\Delta S^{(1)})=0$,
{\it i.e.}, that $\Delta S^{(1)}$  (which
has ghost number zero) should be an observable 
of the undeformed
theory or equivalently
$\Delta S^{(1)}$ is ``BRST-invariant" - 
recall that the solution
$S$ of the master equation generates the BRST 
transformation in the
antibracket.  To
second order in $g$, then, we have
$
(\Delta S^{(1)},\Delta S^{(1)}) + 2
(S, \Delta S^{(2)}) = 0,  \label{obstru}
$
so the antibracket of $\Delta S^{(1)}$ with 
itself should be
the BRST variation of some $\Delta S^{(2)}$.  

Let us start with the full nonlinear 4-dimensional 
$N=1$ case, where a cosmological term {\it can} be
added, for contrast with $D=11$. The action is 
\cite{DZ1}
\be%8
I_4[e^a_\mu, \psi_\lambda] =-\frac{1}{2} \int (dx)
\big(
\frac{1}{2} e e^{a \mu} e^{b \nu} R_{\mu \nu a b}
+
\overline{\psi}_\mu \Gamma^{\mu\sigma\nu} D_\sigma 
\psi_\nu \big),
\label{action40}
\ee
where $e \equiv \det(e_{a \mu})$ and $D_\mu$ here 
is of course
with respect to the full vierbein; it
is invariant  under the local supersymmetry 
(as well as diffeomorphism and local Lorentz)
transformations%
%\footnote{We use first order formalism 
%in $D+4$ and second order in $D=11$, for historical
%reasons only.}
\be%9
\delta e^a_\mu =  - i\bar{\e} \Gamma^a \psi_\m \: ,
\; \;
\delta \psi_\lambda = D_\l \e (x) \: ,
\label{var1}
\ee
and under those of the spin connection 
$\omega^{ab}_\mu$. 
%\; \;
%\delta_\alpha \omega_{a}^{\nu \rho} = e^{-1} 
%\epsilon^{\lambda \mu \nu \rho} \bar{\alpha}
%\gamma_5 \gamma_a D_\lambda \psi_\mu.
%\label{var1}
%\ee
The solution of the master equation takes the 
standard form 
\be
S =  I_4 + \int\int (dx)(dy)\varphi^*_i(x) 
R^i_A (x,y) C^A (y)
 + X,
\label{solmaster1}
\ee
where the $\varphi^*_i$ stand for all the antifields
of antighost number one conjugate to the original
( antighost number zero) fields  $e_{a \mu}$,
$\psi_\lambda$,  and where the
$C^A$ stand for all the ghosts.  The $R^i_A (x,y)$
are the coefficients of all the gauge transformations
leaving $I_4$ invariant.  The terms
denoted by $X$ are at least of antighost number 
two, {\it i.e.}, contain at least two antifields 
$\varphi^*_i$ or one of the antifields
$C^*_\alpha$ conjugate to the ghosts.  The quadratic 
terms in 
$\varphi^*_i$ are also quadratic in the ghosts 
and arise because the gauge transformations do not 
close off-shell \cite{Kallosh}.
We next recall some
cohomological background \cite{HT} related to the 
general solution of the ``cocycle" condition 
$
(S,A) \equiv sA = 0
\label{cocycle}
$
for $A$ with zero ghost number.  If one expands $A$ in
antighost number
$
A = A_0 +  \bar A,
\label{expansion}
$
where $\bar A$ denotes antifield-dependent terms,
one finds that the antifield-independent term 
$A_0$ should be
on-shell gauge-invariant.  Conversely, given an 
on-shell
invariant function(al) $A_0$ of the fields, there is 
a unique, up  to irrelevant ambiguity, solution $A$
(the ``BRST invariant extension" of $A_0$)
that starts with $A_0$.
Below we shall obtain the required $A_0$.
The relevant property that makes the introduction 
of a cosmological
term possible in four dimensions is the fact that 
a gravitino mass term
$ m \int (dx) e \overline{\psi}_\lambda
\Gamma^{\lambda \rho} \psi_\rho \,  $
defines an observable; one easily
verifies that it is on-shell gauge invariant under 
(\ref{var1}).  Hence,
one may complete it with antifield-dependent terms,
to define the initial deformation $m\Delta S^{(1)}$
that satisfies $(\Delta S^{(1)},S) = 0$.
The antifield-dependent contributions are fixed by the
coefficients of the field equations in the gauge variation
of the mass term.  Specifically, since
one must use the {\it undeformed} equations
for the gravitino and the spin connection in order to verify
the invariance of the mass term under supersymmetry
transformations, these contributions will be of the form 
$\psi^* C$ and $\omega^* C$, where
$C$ is the commuting supersymmetry ghost.  They then
lead to the known \cite{Townsend} modification of 
the supersymmetry transformation rules
for the gravitino and the spin connection when the
mass term is turned on\footnote{  A complete
investigation of the BRST cohomology of $N=1$ supergravity 
has been recently
carried out in \cite{Brandt}.}.
Having obtained an acceptable first order deformation,
$m \Delta S^{(1)}$, we must in principle proceed to verify
that $(\Delta S^{(1)}, \Delta S^{(1)})$ is the 
BRST variation of
some $\Delta S^{(2)}$; indeed it is , with $\Delta S^{(2)}=
3/2 \int (dx) e $, as expected.
There are no higher order terms in the deformation
parameter $m$ because
the antibracket of $\Delta S^{(1)}$ with $\Delta 
S^{(2)} $ vanishes ($\Delta S^{(1)}$ does not contain 
the antifields conjugate to the vierbeins), so the 
complete solution of the master equation with 
cosmological constant is 
$
S + m \Delta S^{(1)} + m^2 \Delta S^{(2)}
$, the action of \cite{Townsend,deszum}.
[The possibility of introducing the gravitino mass
term as an observable deformation hinged on the
availability of a dynamical curved geometry in the
sense that while
$(S,\Delta S^{(1)})=0$ is always satisfied, only then is  
$(\Delta S^{(1)}, \Delta S^{(1)})$  BRST exact, i.e.
is there a second order --gravitational-- deformation.]

To summarize the analysis of the four-dimensional 
case, we stress that the cosmological term appears,
in the formulation without auxiliary fields followed here,
as the second order term of a consistent deformation of 
the ordinary supergravity action whose first order 
term is the gravitino mass term, with the
mass as deformation parameter;  
it is completely fixed by the
requirement that the deformation preserve
the master equation and
hence gauge invariance. This means, in particular, that the 
cosmological constant 
itself must be fine-tuned
to the value $-4 m^2$, as  explained 
in \cite{deszum}.\footnote{We emphasize that in
this procedure, one cannot 
start with the cosmological
term as a $\Delta S^{(1)}$.  Indeed, the variation 
of the cosmological term under the gauge 
transformations of the undeformed theory
is algebraic in the fields and hence does not vanish
on-shell, even up to a surface term.  Hence it is not an
observable of the undeformed theory, and so cannot be a
starting point for a consistent deformation:
adding the
cosmological term
(or the sum of it and the mass term) as a $\Delta S^{(1)}$
to the ordinary supergravity action 
is a much more radical (indeed inconsistent !) change
than the gravitino mass term alone.} 
%The cosmological term can only arise as second 
%order term, which explains, as we mentioned above, the
%fact that it is fine-tuned to the value $-m^2$ 
%in terms of the deformation parameter $m$.}

Let us now turn to the action $I_{CJS}$ of \cite{001} in
$D=11$.
The solution of the master equation again takes the standard
form\footnote{Many of the features  of (\ref{solmaster2}) were 
anticipated in \cite{deWit}.}
\be
S =  I_{CJS} 
+ \int \int (dx)(dy)\varphi^*_i(x) R^i_A (x,y) C^A (y)
+ \int (dx)C^{*\mu \nu} \partial_\mu \eta_\nu 
+ \int (dx)\eta^{* \mu} \partial_\mu \rho   + Z,
\label{solmaster2}
\ee
where the $\eta_\nu$ and $\rho$ are the ghosts 
of ghosts and ghost 
of ghost of ghost necessary to account for the 
gauge symmetries of
the $3$-form $A_{\lambda \mu \nu}$, and where $Z$ 
(like $X$ in (6)) is  determined
from the  terms written by the  $(S,S) = 0$ 
requirement.
As in $D=4$, we seek a first-order 
deformation analogous to 
\be
 \Delta S^{(1)} =   \frac{1}{2} m\int (dx) e 
\overline{\psi}_\lambda
\Gamma^{\lambda \rho}
\psi_\rho \,   +
\hbox{ antifield-dep.} \label{mass2}
\ee
However, contrary to what happened at $D=4$, the
mass term no longer defines an observable, as
its variation under local supersymmetry
transformations reads
\be
\delta \big( e \overline{\psi}_\lambda
\Gamma^{\lambda \rho}
\psi_\rho \big) 
\approx  -\frac{i}{18}\overline{\psi}_\mu \Gamma^{\mu
\alpha \beta \gamma \delta} \e 
F_{\alpha \beta \gamma \delta} + O(\psi^3) 
\label{fail}
\ee
where $\approx$ means equal on shell up to a divergence.
%The right-hand side of
%(\ref{fail}) does not vanish on-shell, even up to a total 
%time derivative.
Indeed, the condition that the r.h.s. of (\ref{fail})
 also weakly vanish 
is easily verified to imply, upon 
expansion in the
derivatives of the gauge parameter $\e$, 
that $\overline{\psi_\mu} \Gamma^{\mu\alpha 
\beta \gamma \delta}
\e F_{\alpha \beta \gamma \delta}$ must
vanish on shell, which it does {\it not} do.

Can one improve the first-order deformation 
(\ref{mass2}) to make it acceptable?  The 
cosmological term
will not help because it does not transform 
into $F$.  The
only possible candidates would be functions 
of the 3-form
field.  In order to define observables, these 
functions must
be invariant under the gauge transformations of the
3-form, at least on-shell and up to a total
derivative.  However, in 11 dimensions, the only such
functions can be redefined so as to be
off-shell (and not just on-shell) gauge invariant,
up to a total derivative.  This follows from an 
argument that closely patterns the
analysis of \cite{pform}, defining the very restricted 
class of on-shell invariant vertices  that cannot  in
general be extended  off-shell. [The above 
result actually justifies 
the non-trivial assumption of 
\cite{sagnotti}, that ``on--'' implies ``off--''.]
Thus, the
available functions of $A$ may be assumed to be
strictly gauge invariant, i.e., to be functions
of the field strength  $F$ (which
eliminates $A^2$; also,  changing the coefficient
of the  Chern-Simons
term in the original action clearly cannot  help).  
But it is easy to see that no
expression in $F$ can cancel the unwanted term in
(\ref{fail}),
because of a mismatch in the number of derivatives.
Hence, there is no way to improve the mass term to turn 
it into an observable
in 11 dimensions.  It is the $A$-field part of the
supersymmetry variation of the gravitino that
is responsible for the failure of the mass term to
be an observable, just as it was also 
responsible for the difficulties described 
in the first, linearized, approach. 
Since the cohomology procedure saves us 
from also seeking modifications
of the transformations rules,
we can conclude that the 
introduction of a cosmological constant is 
obstructed already at the first
step in $D=11$ supergravity from the full theory
end as well.

In our discussion, we have assumed (as in
lower dimensions) both that the limit of a 
vanishing mass $m$ is smooth\footnote{This 
restriction is not necessarily stringent: in 
cosmological $D=10$ supergravity \cite{Roman}, there
is  $m^{-1}$ dependence in a field transformation 
rule, but that is an artefact removable
by  introducing a Stuckelberg compensator.}
 and that the field content remains unchanged
in the cosmological variant. Any ``no-go''  result
is of course no stronger than its assumptions, and
ours are shared by the earlier treatments
\cite{Nahm,sagnotti} that we surveyed. There is
one (modest) loosening that can be shown not to
work either, inspired by a recent reformulation
\cite{Green} of the $D=10$ cosmological model 
\cite{Roman}. The idea is to add a deformation 
involving a nonpropagating field, here the 
11-form $G_{11}\equiv
d A_{10}$, through an addition $\Delta I \sim
\int (dx)  [G_{11}+b \bar\psi
\Gamma^9\psi]^2$. The $A_{10}$-field equation states 
that the dual, $\epsilon^{11} [ G_{11}+b
\bar\psi \Gamma^9\psi]$ is a constant of integration,
say $m$. The resulting supergravity field equations 
look like the ``cosmological'' desired ones. 
However, while this ``dualization'' works for 
lower dimensions, in $D=11$ we are simply
back to the original inconsistent model with
supersymmetry still irremediably lost, as  can be 
also discovered 
--without integrating out-- in the deformation approach.

I will end this account with a rather different set of 
``uniqueness" questions that we are currently attempting to 
settle, but that are considerably more speculative.  Here
the invariants whose existence, or rather absence,
we would like to establish are all the possible infinite 
counterterms in a perturbative loop expansion of the theory.
It is of course well-known that all supergravities in  
D$\geq$4 are power counting nonrenormalizable 
{\it a priori}, since the underlying Einstein models are,
so the question is whether supersymmetry can save the day.
But already for D=4, N=1 it was shown early on 
\cite{00x} that at three loops and higher, 
suitable supersymmetric invariants
existed, and it would be very unlikely if their 
coefficients precisely vanished in (impossible to 
perform!) explicit calculations.  Now strictly 
speaking, before 
talking about candidate terms,
one must first exhibit a regularization scheme that
preserves the supersymmetry, something notoriously
difficult in odd dimensions (due to the Levi--Civita
symbol, for example).  So we cannot point to dimensional
regularization as a legitimate scheme, but let us 
nevertheless carry on formally within it 
and seek terms that are ~a) supersymmetric, ~b)
dimensionally correct in a loop expansion in the 
sole dimensional constant of the theory, the 
gravitational one.
Recall that the Einstein term $\k^{-2}R$ in D=11 fixes
the dimension of $\k^2$ to be $L^9$.  
The constant $\k$ also appears in front of the   
form field's famous
Chern--Simons term, $\kappa \e FFA$ (here parity 
preserving!), as is clear by comparing 
its dimension with the kinetic term $F^2$.

Since already the gravitational parts of local 
counterterms, being of the form $R^n$ (possibly 
involving an even number of
covariant derivatives) are even-dimensional, only odd
powers of $\k^2$ and hence only {\it even} loops can 
contribute to a local integral over 
$(d^{11}x)$.  This ``counting" fact has long been known
({\it e.g.}, \cite{Duff}) although strictly speaking there
exists a gravitational Chern--Simons term of the form
$\e^{1...11}R_1 ..R_5 \omega$ that has odd dimensions.
However, it has odd parity and so should not
arise in this parity even model (unless there are
anomalies).  
Optimistically, then, one need only worry
about 2k-loop invariants, and then indeed only about the
subset of invariants that fail to vanish on-shell; those
that do vanish there can always be absorbed by a harmless 
field-redefinition \cite{00y}.  The simplest, two-loop,
contribution would presumably begin as 
$\k^{+2} \int d^{11}x \, \D L_2$, with the leading
gravitational parts $\D L_2 \sim R^{10} + (DR)^2R^7
+ .. (D^8R)^2$ in a very schematic notation; the 
$R$'s are all Weyl tensors and $D$ represents a
covariant derivative.  Likewise the $F$-field would
enter through invariants of suitable powers of $F$ and
their derivatives, in addition to dimensionally relevant
fermionic and mixed
terms.  To test the hypothesis that 
(as in the cosmological case) it is
the $F$ field that is the culprit, one can begin with
candidate polynomials in $F$ alone and vary them, looking
for obstructions to supersymmetry.  There are 
some indications that such obstructions are present, 
but we don't yet have a systematic way to classify:
the simple (?)  combinatorial preliminary,
exhibiting the local invariants that
can be constructed from a 4-form in D=11, is
not yet systematically known.  The idea of our 
procedure is that the supersymmetric variations of some
initial $F^n$ term is $\sim F^{n-1} \bar{\a} \G f$
where $f_{\m\n} \equiv
D_\m \psi_\n -D_\n \psi_\m$ is the fermionic 
field strength.  To cancel this
variation requires a companion term $\sim \bar{f}
\G f \; F^{n-2}$, which will in turn also vary into
gravity, and one may hope that -- as we saw with our
cosmological construction --  the process cannot be
completed. Although the above idea may not 
be easy to test
without some deeper understanding of the theory, it
is bound to teach us more about this one QFT
that survives at our present level of
post-string unification!

\bigskip
The work of S. D. and D. S. was supported by  
NSF grant PHY 93-15811

\end{document}